\documentclass[prd,preprint,superscriptaddress,showpacs,byrevtex]{revtex4} 
\usepackage{epsfig} 
\newcommand{\be}{\begin{equation}} 
\newcommand{\ee}{\end{equation}} 
\newcommand{\bea}{\begin{eqnarray}} 
\newcommand{\eea}{\end{eqnarray}} 

\begin{document} 

\begin{flushright}
YITP-SB-06-49
\end{flushright}

\title{\noindent Schwinger Mechanism in the Presence of Arbitrary Time 
Dependent Background Electric Field }

\author{Fred Cooper} \email{fcooper@nsf.gov} 
\affiliation{Physics Division, National Science Foundation, Arlington VA 22230}

\author{Gouranga C. Nayak} \email{nayak@insti.physics.sunysb.edu} 
\affiliation{ C. N. Yang Institute for Theoretical Physics, Stony Brook University, SUNY, Stony Brook, 
NY 11794-3840, USA } 

\date{\today} 

\begin{abstract} 

We study, for the first time, the Schwinger mechanism for the pair production
of charged scalars in the presence of an arbitrary time-dependent  
background electric field $E(t)$ by by directly evaluating the path
integral. We obtain an exact non-perturbative result for the probability 
of charged scalar particle-antiparticle pair production per unit 
time per unit volume per unit transverse momentum 
(of the particle or antiparticle) from the arbitrary time dependent electric 
field $E(t)$. We find that the exact non-perturbative 
result is independent of all the time derivatives $\frac{d^nE(t)}{dt^n}$, 
where $n=1,2,....\infty$. This result has the same functional dependence on $E$
as the constant  electric field $E$ result with the replacement: 
$E~\rightarrow~ E(t)$.

\end{abstract} 

\pacs{PACS: 11.15.-q, 11.15.Me, 11.15.Tk, 11.15.-z} %

\maketitle 

\newpage 
In a classic paper in 1951 Schwinger derived the following one-loop non-perturbative formula
\bea
\frac{dW}{d^4x}=\frac{e^2E^2}{4\pi^3} ~\sum_{n=1}^{\infty}  ~\frac{1}{n^2}~e^{-\frac{n\pi m^2}{|eE|}}
\eea
for the probability of $e^+e^-$ pair production per unit time per unit volume 
from a constant electric field E via vacuum polarization \cite{schw} by
using proper time method. In case of charged 
scalar field theory the corresponding result is given by
\bea
\frac{dW}{d^4x}=\frac{e^2E^2}{8\pi^3} ~\sum_{n=1}^{\infty} ~\frac{(-1)^{n+1}}{n^2}~e^{-\frac{n\pi m^2}{|eE|}}.
\eea
The $p_T$ distribution of the $e^+$ (or $e^-$) production, 
$\frac{dW}{d^4x d^2p_T}$, can not be obtained by using proper time method and a WKB approximate
method was used for this purpose in \cite{Casher}. A path integral method
can also be used to obtain the $p_T$ distribution \cite{gouranga}. In the case 
of fermions in QED one finds \cite{Casher}
\bea
\frac{dW}{d^4xd^2p_T}=-\frac{|eE|}{4\pi^3} {\rm Log}[1-e^{-\pi \frac{p_T^2+m^2}{|eE|}}].
\label{dsf}
\eea
The corresponding result for the charged scalar production is given by
\bea
\frac{dW}{d^4xd^2p_T}=\frac{|eE|}{8\pi^3} {\rm Log}[1+e^{-\pi \frac{p_T^2+m^2}{|eE|}}].
\label{ds}
\eea

Recently  this approach has been extended to QCD in SU(3)
gauge group by directly evaluating the path integral and the $p_T$ distribution of the quark and gluon 
production rate are found to depend on two independent Casimir invariants: $C_1=[E^aE^a]$ and
$C_2=[d_{abc}E^aE^bE^c]^2$ where $E^a$ is the constant chromo-electric field with color
index a=1,2,..8 \cite{gouranga}. 

In this paper we study, for the first time, Schwinger mechanism in the presence of  an arbitrary time 
dependent electric field $E(t)$ by directly evaluating the path integral.  This is possible by using
recently derived shift theorems for path integrals \cite{shift} and the fact that the eigenvectors of the
Hamiltonian for a harmonic oscillator with time dependent frequency 
$\omega (t)$ are the {\it same} as the
eigenvectors of the constant frequency $\omega$ 
problem, with the replacement $\omega \rightarrow \omega(t)$ \cite{oscillator1,oscillator2}.

Our result which we will derive below is  the following non-perturbative 
formula for the probability of charged scalar particle-antiparticle 
pair production per unit time per unit volume per unit transverse momentum (of the particle or antiparticle)
production from arbitrary time dependent electric field $E(t)$ via Schwinger 
mechanism 
\bea
\frac{dW}{d^4xd^2p_T}=\frac{|eE(t)|}{8\pi^3} {\rm Log}[1+e^{-\pi \frac{p_T^2+m^2}{|eE(t)|}}].
\label{dw}
\eea

This result has the remarkable feature that no time derivatives 
$\frac{d^nE(t)}{dt^n}$ (where n=1,2,...$\infty$)
occur in the final expression. This implies that the slowly varying
(adiabatic) approximation \cite{cooper}, 
where one merely replace $E$ by $E(t)$ 
in the formulas for pair production appears to be exact
after Bose Enhancement effects are included. 

Now we present a derivation of eq. (\ref{dw}).

The vacuum to vacuum transition amplitude in the presence of 
background field $A$ is given by
\bea
<0|0>^A =\frac{\int [d \phi^\dag][d\phi] e^{i\int d^4x \phi^\dag M[A] \phi}}{\int[d \phi^\dag][d\phi] e^{i\int d^4x \phi^\dag M[0] \phi}}
=Det^{-1}[M[A]]/Det^{-1} [M[0]] =e^{iS^{(1)}}
\label{1}
\eea
where for scalar field theory
\bea
M[A]=({\hat p}-eA)^2-m^2,  ~~~~~~{\rm with} ~~~~~~~{\hat p}_\mu=\frac{1}{i}\frac{\partial}{\partial x^\mu}.
\label{2}
\eea
The one loop effective action is
\bea
S^{(1)}=i{\rm Tr ln} [({\hat p}-eA)^2 -m^2] -i {\rm Tr ln} [\hat{p}^2  -m^2],
\label{3}
\eea
where the ${\rm Tr}$ is given by
\bea
{\rm Tr} {\cal O}=\int d^4x <x| {\cal O} |x>.
\label{4}
\eea
Since it is convenient to work with the trace of the exponential
we replace the logarithm by 
\bea
{\rm ln} \frac{a}{b}=\int_0^\infty \frac{ds}{s} [
e^{-is(b-i\epsilon)} -e^{-is(a-i\epsilon)}].
\label{5}
\eea
Hence we get from eq. (\ref{3})
\bea
S^{(1)}=-i\int_0^\infty \frac{ds}{s} {\rm Tr} [
e^{-is[(\hat{p}-eA)^2-m^2-i\epsilon]} -e^{-is(\hat{p}^2-m^2-i\epsilon)}].
\label{6}
\eea
We assume that the time dependent electric field E(t)
is along the z-axis. We choose the Axial gauge $A_3=0$  
so that  only
\bea
A_0 =- E(t)z.
\label{7}
\eea
is non-vanishing. Using eq. (\ref{7}) in eq. (\ref{6}) we obtain
\bea
&& S^{(1)}=-i \int_0^\infty \frac{ds}{s} 
\int_{-\infty}^{+\infty} dt <t| \int_{-\infty}^{+\infty} dx 
<x| \int_{-\infty}^{+\infty} dy <y| \int_{-\infty}^{+\infty} dz <z| \nonumber \\
&& [ e^{-is[(\hat{p}_0+eE(t) z)^2-\hat{p}_z^2-\hat{p}_T^2-m^2-i\epsilon]} -
e^{-is(\hat{p}^2-m^2-i\epsilon)}] |z> |y> |x> |t>.
\label{10}
\eea
Inserting complete set of $|p_T>$ states $\int d^2 p_T |p_T><p_T|~=~1$ 
we find (we use the normalization $<q|p>=\frac{1}{\sqrt{2\pi}} ~e^{iqp}$)
\bea
&& S^{(1)}=\frac{-i }{(2\pi)^2}
\int_0^\infty \frac{ds}{s} \int d^2x_T \int d^2p_T
e^{is(p_T^2+m^2+i\epsilon)}[\int_{-\infty}^{+\infty} dt <t|  
\int_{-\infty}^{+\infty} dz <z| \nonumber \\
&& e^{-is[(-i\frac{d}{dt}+eE(t) z)^2-\hat{p}_z^2]} |z> |t>- \int dt \int dz 
\frac{1}{4\pi s}].
\label{11}
\eea
At this stage we use the shift theorem \cite{shift} 
\bea
&& \int dy \int_{-\infty}^{+\infty} dx f_1(y)<x|~e^{-[(a(y)x+h\frac{d}{dy})^2+b(\frac{d}{dx})+
c(y)]}
~|x>f_2(y) \nonumber \\
&&= \int dy \int_{-\infty}^{+\infty} dx f_1(y)<x
-\frac{h}{a(y)}\frac{d}{dy}|~e^{-[a^2(y)x^2+b(\frac{d}{dx})+c(y)]} ~|x
-\frac{h}{a(y)}\frac{d}{dy}>f_2(y) \nonumber \\
\eea
to obtain:
\bea
&& S^{(1)}=\frac{i}{(2\pi)^2} \int_0^\infty\frac{ds}{s} \int d^2x_T\int d^2p_T
e^{is(p_T^2+m^2+i\epsilon)} \nonumber \\
&& [ \int dt \int dz \frac{1}{4\pi s}- \int_{-\infty}^{+\infty} dt <t|
\int_{-\infty}^{+\infty} dz <z+\frac{i}{E(t)}\frac{d}{dt}| 
 e^{-is[e^2E^2(t)z^2-\hat{p}_z^2]}|z+\frac{i}{E(t)}\frac{d}{dt}>|t>]\nonumber \\
\label{12a}
\eea
where the $z$ integration must be performed from $-\infty$ to $+\infty$
for the shift theorem to be applicable \cite{shift}.

Inserting complete set of $|p_z>$ states (using $\int dp_z |p_z><p_z|=1$)
as appropriate we find
\bea
&& S^{(1)}=\frac{i}{(2\pi)^2} \int_0^\infty \frac{ds}{s} \int d^2x_T \int d^2p_T
e^{is(p_T^2+m^2+i\epsilon)} \nonumber \\
&&[ \int dt \int dz ~\frac{1}{4\pi s}- \int_{-\infty}^{+\infty} dt <t|\int dp_z\int dq_z  
\int_{-\infty}^{+\infty} dz <z+\frac{i}{E(t)}\frac{d}{dt}|p_z> \nonumber \\
&& <p_z| e^{is[-e^2E^2(t) z^2+\hat{p}_z^2]}|q_z> <q_z|z+\frac{i}{E(t)}\frac{d}{dt}>|t>] \nonumber \\
&& =\frac{i}{(2\pi)^2} \int_0^\infty \frac{ds}{s} \int d^2x_T\int d^2p_T
e^{is(p_T^2+m^2+i\epsilon)} [ \int dt \int dz~ \frac{1}{4 \pi s}- F]
\label{12}
\eea
where
\bea
&& F=\frac{1}{(2\pi)} \int_{-\infty}^{+\infty} dt <t|\int dp_z\int dq_z
\int_{-\infty}^{+\infty} dz~ e^{izp_z}e^{-\frac{1}{E(t)}\frac{d}{dt}p_z}<p_z| 
 e^{is[-e^2E^2(t) z^2+\hat{p}_z^2]}|q_z>  \nonumber \\
 && e^{\frac{1}{E(t)}\frac{d}{dt} q_z}
e^{-izq_z}|t>.
\label{gn}
\eea
It can be seen that the exponential $e^{-\frac{1}{E(t)}\frac{d}{dt}p_z}$ 
contains the derivative $\frac{d}{dt}$ which operates on
$<p_z|e^{is[-e^2E^2(t) z^2+\hat{p}_z^2]}|q_z> e^{\frac{1}{E(t)}\frac{d}{dt} q_z}$ 
hence we can not move $e^{-\frac{1}{E(t)}\frac{d}{dt}p_z}$ to right.

We insert complete sets of $|z>$ states  
and $|p_0> $states using the completeness relations:
\be
1= \int_{-\infty}^ {\infty} dz |z> < z|; ~~~    1= \int_{-\infty}^ {\infty} dp_0 |p_0> < p_0|
\ee
to find
\bea
&& F=\frac{1}{(2\pi)} \int_{-\infty}^{+\infty} dt \int dp_0 \int dp'_0 \int dp''_0 \int dp'''_0 \int dp_z\int dq_z \int dz_1 \int dz_2
\int_{-\infty}^{+\infty} dz \nonumber \\
&& <t|p_0><p_0|e^{izp_z}e^{-\frac{1}{E(t)}\frac{d}{dt}p_z}|p'_0><p'_0|<p_z|z_1> 
<z_1| e^{is[-e^2E^2(t) z^2+\hat{p}_z^2]}|z_2><z_2|q_z>|p''_0> \nonumber \\
&& <p''_0|e^{\frac{1}{E(t)}\frac{d}{dt} q_z} e^{-izq_z}|p'''_0><p'''_0|t> \nonumber \\
&& =\frac{1}{(2\pi)^3} \int_{-\infty}^{+\infty} dt \int dp_0 \int dp'_0 \int dp''_0 \int dp'''_0 \int dp_z\int dq_z \int dz_1 \int dz_2
\int_{-\infty}^{+\infty} dz \nonumber \\
&& e^{itp_0} e^{izp_z}<p_0|e^{-\frac{1}{E(t)}\frac{d}{dt}p_z}|p'_0>e^{-iz_1p_z} <p'_0|
<z_1| e^{is[-e^2E^2(t) z^2+\hat{p}_z^2]}|z_2> |p''_0> \nonumber \\
&& e^{iz_2q_z} <p''_0|e^{\frac{1}{E(t)}\frac{d}{dt} q_z} |p'''_0> e^{-izq_z}e^{-itp'''_0}.
\label{gn5}
\eea

The matrix element, $<z_1| e^{is[-e^2E^2(t) z^2+\hat{p}_z^2]}|z_2>$ 
can be evaluated in terms of the eigenstates of the Harmonic Oscillator 
with imaginary \cite{itzy,gouranga} and time dependent frequency
$\omega(t)= ieE(t)$. The Hamiltonian 
\be
H= p_z^2 + \omega^2(t) z^2
\ee
can be diagonalized at every time $t$ in terms of 
squeezed state time dependent creation and 
annihilation operators $b^\dag(t)$ and $b(t)$ 
\cite{oscillator2}. Explicitly
\be
H = \omega(t) [ 2 b^\dag(t) b(t) + 1]. 
\ee
The eigenvalues of the Hamiltonian are just
\be
\lambda_n = \omega(t) [2 n+1], ~~~~n= 0, 1,2, \cdots.
\ee
and the normalized eigenstates  $<z|n_t>$  are the usual harmonic oscillator ones with $\omega \rightarrow \omega(t)$.

\bea
[{\hat p}_z^2 + \omega^2(t)z^2]|n_t>=(2n+1)\omega(t) |n_t>
\eea
\bea
<z|n_t>=\psi_n(z)=(\frac{\omega(t)}{\pi})^{1/4}\frac{1}{(2^nn!)^{1/2}}
H_n(z \sqrt{\omega(t)})e^{-\frac{\omega(t)}{2}z^2}.
\label{hmt}
\eea
Note that the above "instantaneous" wave functions with the eigen
values $[(2n+1) \omega(t)]$ can also
be directly obtained by constructing creation and anihillation
operators from $z$, $p_z$ and $\omega(t)$ (like constant $\omega$ problem)
irrespective of the works in \cite{oscillator1,oscillator2}. 
By using the orthonormality relation of Hermite polynomials we find
\bea
\int dz  |\psi_n(z)|^2=\int dz |<n_t|z>|^2 = \int dz (
\frac{\omega(t)}{\pi})^{1/2}\frac{1}{(2^nn!)}|H_n(z 
\sqrt{\omega(t)})|^2e^{-\omega(t)z^2}=1
\label{nor}
\eea
and
\bea
&& \int dz \psi^{*}_n(z) \psi_m(z) = \int dz <n_t|z><z|m_t> \nonumber \\
&& = \int dz (\frac{\omega(t)}{\pi})^{1/2}\frac{1}{(2^nn!)} 
H^{*}_n(z \sqrt{\omega(t)})H_m(z \sqrt{\omega(t)})e^{-\omega(t)z^2}
=\delta_{nm}.
\label{orth}
\eea
By using $\int dz |z><z| =1$ in the above equation we find
\bea
<n_t|m_t>=\delta_{nm}.
\label{o1}
\eea
The completeness condition is
\bea
\sum_n |n_t><n_t|=1. 
\label{c1}
\eea
Inserting complete set of harmonic oscillator states $|n_t>$
( $\sum_n |n_t><n_t|=1$) in eq. (\ref{gn5}) we find
\bea
&& F=
\frac{1}{(2\pi)^3} 
\sum_n \sum_m \int_{-\infty}^{+\infty} dt \int dp_0 \int dp'_0 \int dp''_0 \int dp'''_0 \int dp_z\int dq_z \int dz_1 \int dz_2
\int_{-\infty}^{+\infty} dz \nonumber \\
&& e^{itp_0} e^{izp_z}<p_0|e^{-\frac{1}{E(t)}\frac{d}{dt}p_z}|p'_0>e^{-iz_1p_z} <p'_0|
<z_1|n_t> <n_t|e^{is[-e^2E^2(t) z^2+\hat{p}_z^2]}|m_t> \nonumber \\
&& <m_t|z_2> |p''_0> e^{iz_2q_z} <p''_0|e^{\frac{1}{E(t)}\frac{d}{dt} q_z} |p'''_0> e^{-izq_z}e^{-itp'''_0} \nonumber \\
&& =
\frac{1}{(2\pi)^3} 
\sum_n  \int_{-\infty}^{+\infty} dt \int dp_0 \int dp'_0 \int dp''_0 \int dp'''_0 \int dp_z\int dq_z \int dz_1 \int dz_2
\int_{-\infty}^{+\infty} dz \nonumber \\
&& e^{itp_0} e^{izp_z}<p_0|e^{-\frac{1}{E(t)}\frac{d}{dt}p_z}|p'_0>e^{-iz_1p_z} <p'_0|
<z_1|n_t> e^{-seE(t)(2n+1)}<n_t|z_2> |p''_0> \nonumber \\
&& e^{iz_2q_z} <p''_0|e^{\frac{1}{E(t)}\frac{d}{dt} q_z} |p'''_0> e^{-izq_z}e^{-itp'''_0} \nonumber \\
&& =
\frac{1}{(2\pi)^3} 
\sum_n  \int_{-\infty}^{+\infty} dt \int dp_0 \int dp'_0 \int dp''_0 \int dp'''_0 \int dp_z\int dq_z \int dz_1 \int dz_2
\int_{-\infty}^{+\infty} dz \nonumber \\
&& e^{itp_0} e^{iz(p_z-q_z)}<p_0|e^{-\frac{1}{E(t)}\frac{d}{dt}p_z}|p'_0>e^{-iz_1p_z} <p'_0|
<z_1|n_t> e^{-seE(t) (2n+1)}<n_t|z_2> |p''_0> \nonumber \\
&& e^{iz_2q_z} <p''_0|e^{\frac{1}{E(t)}\frac{d}{dt} q_z} |p'''_0> e^{-itp'''_0}.
\label{gn7}
\eea
As advocated earlier 
the $z$ integration must be performed from $-\infty$ to $+\infty$
for the shift theorem to be applicable \cite{shift}.
Performing  the $z$ integration we get
 \bea
\int_{-\infty}^{+\infty} dz e^{iz(p_z-q_z)} = 2 \pi \delta(p_z-q_z)
\eea
and we find
\bea
&&F =
\frac{1}{(2\pi)^2} 
\sum_n  \int_{-\infty}^{+\infty} dt \int dp_0 \int dp'_0 \int dp''_0 \int dp'''_0 \int dp_z \int dz_1 \int dz_2 e^{itp_0} 
<p_0|e^{-\frac{1}{E(t)}\frac{d}{dt}p_z}|p'_0> \nonumber \\
&& e^{-iz_1p_z} <p'_0| <z_1|n_t> e^{-seE(t) (2n+1)} <n_t|z_2> |p''_0> e^{iz_2p_z} <p''_0|e^{\frac{1}{E(t)}\frac{d}{dt} p_z} |p'''_0> e^{-itp'''_0}. \nonumber \\
\label{gn8}
\eea
Inserting further complete sets of $|t>$ states we obtain
\bea
&&F =
\frac{1}{(2\pi)^2} 
\sum_n  \int_{-\infty}^{+\infty} dt \int dt_1 \int dt_2 
\int dp_0 \int dp'_0 \int dp''_0 \int dp'''_0 \int dp_z \int dz_1 \int dz_2 
 e^{itp_0} \nonumber \\
 && <p_0|e^{-\frac{1}{E(t)}\frac{d}{dt}p_z}|p'_0> e^{-iz_1p_z} <p'_0|t_1>
<t_1|<z_1|n_t> e^{-seE(t) (2n+1)}<n_t|z_2> \nonumber \\
&& |t_2><t_2|p''_0>  e^{iz_2p_z} <p''_0|e^{\frac{1}{E(t)}\frac{d}{dt} p_z} |p'''_0> e^{-itp'''_0} \nonumber \\
&& =
\frac{1}{(2\pi)^3} 
\sum_n  \int_{-\infty}^{+\infty} dt \int dt_1 \int dt_2 
\int dp_0 \int dp'_0 \int dp''_0 \int dp'''_0 \int dp_z \int dz_1 \int dz_2 \nonumber \\
&& e^{itp_0} <p_0|e^{-\frac{1}{E(t)}\frac{d}{dt}p_z}|p'_0>e^{-iz_1p_z} e^{-it_1p'_0}
<t_1|<z_1|n_t> e^{-seE(t) (2n+1)}<n_t|z_2> \nonumber \\
&& |t_2>e^{it_2p''_0} e^{iz_2p_z} <p''_0|e^{\frac{1}{E(t)}\frac{d}{dt} p_z} |p'''_0> e^{-itp'''_0}.
\label{gn9}
\eea
 Since $<n_t|z>$ does not contain any time derivative $\frac{d}{dt}$ operator 
 (see eq. (\ref{hmt})) we find (by using $<t_1|t_2>=\delta(t_1-t_2)$)
 \bea
&&F =
\frac{1}{(2\pi)^3} 
\sum_n  \int_{-\infty}^{+\infty} dt \int dt_1 
\int dp_0 \int dp'_0 \int dp''_0 \int dp'''_0 \int dp_z \int dz_1 \int dz_2 \nonumber \\
&& e^{itp_0} <p_0|e^{-\frac{1}{E(t)}\frac{d}{dt}p_z}|p'_0>e^{-iz_1p_z} e^{-it_1p'_0}
<z_1|n_{t_1}> e^{-seE(t_1) (2n+1)}<n_{t_1}|z_2> e^{it_1p''_0} \nonumber \\
&& e^{iz_2p_z} <p''_0|e^{\frac{1}{E(t)}\frac{d}{dt} p_z} |p'''_0> e^{-itp'''_0}.
\label{gn10}
\eea
Since $<p_0|e^{-\frac{1}{E(t)}\frac{d}{dt}p_z}|p'_0>$ and
$<p''_0|e^{\frac{1}{E(t)}\frac{d}{dt} p_z} |p'''_0>$ are independent of the time derivative
operator $\frac{d}{dt}$ (they only depend on c-numbers $p_0$, $p'_0$, $p''_0$ and $p'''_0$)
we can take $<p''_0|e^{\frac{1}{E(t)}\frac{d}{dt} p_z} |p'''_0>$ and $e^{-itp'''_0}$ to the left. We find
 \bea
&&F =
\frac{1}{(2\pi)^3} 
\sum_n  \int_{-\infty}^{+\infty} dt \int dt_1 
\int dp_0 \int dp'_0 \int dp''_0 \int dp'''_0 \int dp_z \int dz_1 \int dz_2 \nonumber \\
&& e^{it(p_0-p'''_0)} <p''_0|e^{\frac{1}{E(t)}\frac{d}{dt} p_z} |p'''_0><p_0|e^{-\frac{1}{E(t)}\frac{d}{dt}p_z}|p'_0>e^{-iz_1p_z} e^{-it_1p'_0}
<z_1|n_{t_1}> e^{-seE(t_1) (2n+1)} \nonumber \\
&& <n_{t_1}|z_2> e^{it_1p''_0} e^{iz_2p_z}.
\label{gn12}
\eea
It can also be shown that $<p_0|e^{-\frac{1}{E(t)}\frac{d}{dt}p_z}|p'_0>$ and
$<p''_0|e^{\frac{1}{E(t)}\frac{d}{dt} p_z} |p'''_0>$ are independent of
$t$. This can be shown as follows:
\bea
&& <p_0|f(t)\frac{d}{dt}|p'_0>= \int dt' \int dt'' \int dp''''_0 <p_0|t'> <t'|f(t)|t''><t''|p''''_0><p''''_0|\frac{d}{dt}|p'_0> \nonumber \\
&&= \int dt' \int dt'' \int dp''''_0 e^{-it' p_0}~ \delta(t'-t'') f(t'')
e^{it'' p''''_0}~
ip'_0~ \delta(p''''_0-p'_0)=ip'_0 \int dt'  ~e^{-it'(p_0-p'_0)}f(t') \nonumber \\
\eea
which is independent of $t$. Hence we can easily integrate over $t$ in eq.
(\ref{gn12}).

Integrating over $t$ in eq. (\ref{gn12}) by using $\int_{-\infty}^{+\infty} dt e^{it(p_0-p'''_0)} = 2 \pi \delta(p_0-p'''_0)$, we find
 \bea
&&F =
\frac{1}{(2\pi)^2} 
\sum_n  \int dt_1 
\int dp_0 \int dp'_0 \int dp''_0  \int dp_z \int dz_1 \int dz_2 <p''_0|e^{\frac{1}{E(t)}\frac{d}{dt} p_z} |p_0> <p_0|e^{-\frac{1}{E(t)}\frac{d}{dt}p_z}|p'_0> \nonumber \\
&& e^{-iz_1p_z} e^{-it_1p'_0}
<z_1|n_{t_1}> e^{-seE(t_1) (2n+1)} <n_{t_1}|z_2> e^{it_1p''_0} e^{iz_2p_z}.
\label{gn13}
\eea
Since $\int dp_0 |p_0><p_0| =1$ and 
$\int dp_z e^{ip_z(z_1-z_2)}=2\pi \delta(z_1-z_2)$ we find
\bea
&&F=
\frac{1}{(2\pi)^2} 
\sum_n  \int dt_1 
 \int dp'_0 \int dp_z \int dz_1 \int dz_2  e^{-iz_1p_z}
<z_1|n_{t_1}> e^{-seE(t_1) (2n+1)} <n_{t_1}|z_2>  e^{iz_2p_z} \nonumber \\
&& =
\frac{1}{(2\pi)} 
\sum_n  \int dt  \int dp'_0  \int dz 
|<z|n_{t}>|^2 ~e^{-seE(t) (2n+1)}.
\label{gn15}
\eea
Since the harmonic oscillator wave function is normalized (see eq. (\ref{nor})):
\bea
\int dz |<z|n_t>|^2=1
\eea
we find
\bea
F= \frac{1}{(2\pi)} 
\sum_n  \int dt  \int dp_0 e^{-seE(t) (2n+1)} = \frac{1}{(2\pi)} 
\int dt  \int dp_0 \frac{1}{{2 \rm sinh}(seE(t))}
\label{gn16}
\eea
Using the Lorentz force equation
\bea
dp_\mu = eF_{\mu \nu}dx^\nu
\eea
we find (when $E(t)$ is along $z$-axis, eq. (\ref{7}))
\bea
dp_0=eE(t) dz.
\label{dt}
\eea
Hence 
\bea
F=
\frac{1}{(4\pi)} 
\int dt  \int dz  \frac{eE(t)}{{ \rm sinh}(seE(t))}.
\label{gn17}
\eea
Using the above expression for $F$ in eq. (\ref{12}) the effective action becomes
\bea
S^{(1)}=
\frac{i}{16\pi^3} 
\int_0^\infty \frac{ds}{s} \int d^4x \int d^2p_T
e^{is(p_T^2+m^2+i\epsilon)} 
[ \frac{1}{ s}- \frac{eE(t)}{{\rm sinh}(seE(t))}]. 
\label{12ff}
\eea
The series expansion for $\frac{1}{{\rm sinh} x}$ is given by
\bea
\frac{1}{{\rm sinh} x}=\frac{1}{x} + 2x \sum_{n=1}^\infty \frac{(-1)^n}{\pi^2 n^2 +x^2}.
\eea
Hence
\bea
S^{(1)}=
\frac{-i}{16\pi^3} 
\sum_{n=1}^\infty \int_0^\infty \frac{ds}{s} \int d^4x \int d^2p_T eE(t)
e^{is(p_T^2+m^2+i\epsilon)} \frac{(-1)^n 2seE(t)}{n^2\pi^2+s^2e^2E^2(t)}. 
\label{13ff}
\eea
The s-contour integration is straight forward \cite{gouranga,itzy,schw}. 
Performing the s-contour integration around the pole $s=\frac{in\pi}{|eE(t)|}$
we find for the probability  $W$ of pair production (which is twice
the  imaginary part of the effective action)
\bea
W=2 {\rm Im} S^{(1)}=
\frac{1}{8\pi^3} 
\sum_{n=1}^\infty \frac{(-1)^{n+1}}{n} \int d^4x \int d^2p_T |eE(t)|
e^{-n\pi \frac{p_T^2+m^2}{|eE(t)|}}.
\label{14ff}
\eea
Therefore the probability of producing a charged particle per unit volume 
per unit time with transverse momentum $p_T$ is given by
\bea
\frac{dW}{d^4xd^2p_T}=
\frac{|eE(t)|}{8\pi^3} {\rm Log}[1+e^{-\pi \frac{p_T^2+m^2}{|eE(t)|}}]
\label{dwf}
\eea
which reproduces eq. (\ref{dw}).

To conclude we have studied, for the first time,
the Schwinger mechanism in the presence
of arbitrary time dependent background electric field E(t)
by directly evaluating the path integral. We have obtained an exact one-loop non-perturbative 
result for the probability of charged scalar particle-antiparticle pair production per unit time per unit 
volume per unit transverse momentum (of the particle or antiparticle)
from the arbitrary time dependent electric field $E(t)$. We have 
found that the exact non-perturbative result is independent of all the time derivatives $\frac{d^nE(t)}{dt^n}$, where $n=1,2,....\infty$. 

The quantum Back Reaction problem for the Electric Field is studied in 
\cite{cooper}.  For the Back-Reaction problem, the equation for the charged scalar field  is supplemented by the Back-Reaction equation for the time evolution of the Electric Field , which is the semiclassical
Maxwell's equation:
\be
\frac {\partial E(t)}{\partial t} = - \langle  j^3 (t) \rangle
\ee
where  $j^3(t)$ is the induced (quantum) current coming from the produced  charged scalar
mesons, and the expectation value is taken with respect to the initial adiabatic vacuum state of the 
scalar field theory.   The exact numerical solution of the
coupled semiclassical Maxwell equation and the equation for the time evolution of the quantum field $\phi$ were compared in those papers to the solution of 
of the semi-classical transport equation
having a source term for pair production which utilizes Schwinger's
constant electric field $E$  result but with the replacement 
$E~\rightarrow E(t)$.   The transport approach agreed with a  coarse graining of the quantum field theory
solution. 
This result  seems to be consistent
with our finding here. However in those simulations the Schwinger source term was modified by
Bose Enhancement effects for boson pair production and Pauli-blocking effects for fermion pair production \cite{cooper,tr1}. These enhancements or (Pauli-blocking) effects are not present in our calculation and need to be better understood.
Particle production from arbitrary time (and space) dependent background 
fields is important in early universe dynamics  and in the  production and 
equilibration of the quark-gluon plasma at RHIC and LHC \cite{cooper,tr1,grr1}. 

\acknowledgments
We thank George Sterman and Peter van Nieuwenhuizen for 
discussions. This work was supported in part by the National Science 
Foundation, grants PHY-0354776 and PHY-0345822. Fred Cooper would like 
to thank Harvard University for its hospitality during the writing of 
this paper.


\begin{thebibliography}{99}
\bibitem{schw} J. Schwinger, Phys. Rev. 82 (1951) 664.
\bibitem{Casher} A. Casher, H. Neuberger, and S. Nussinov, Phys. Rev. D 20, 179 (1979).
\bibitem{gouranga}  G. Nayak and P. van Nieuwenhuizen  Phys. Rev. D71 (2005) 125001; F. Cooper and G. C. Nayak, Phys. Rev. D73 (2006) 065005;
G. C. Nayak, Phys. Rev. D72 (2005) 0510052.
\bibitem{shift} F. Cooper and G. C. Nayak, hep-th/0609192.
\bibitem{oscillator1} H. R. Lewis, Jr. and W. B. Reisenfeld, J. Math. Phys. 10, (1969) 1458.
\bibitem{oscillator2} K. Yeon, H. Kim, . Um, T. Gorge and L. Pandey,  Phys. Rev. A50 (1994) 1035.
\bibitem{itzy} C. Itzykson and J-B. Zuber, Quantum Field Theory, page-194, 
Dover Publication, Inc. Mineola, New York.
\bibitem{cooper} F. Cooper and E. Mottola, Phys. Rev. D {\bf 40}, 456 (1989);
Y. Kluger, J. M. Eisenberg, B. Svetitsky, F.  Cooper and E. Mottola, 
Phys. Rev. Lett. 67 (1991) 2427;
Y. Kluger, J. M. Eisenberg, B. Svetitsky, F.  Cooper and E. Mottola,
 Phys. Rev. D {\bf 45} (1992)4659. F. Cooper, J.M. Eisenberg, Y. Kluger, E.
Mottola, and B. Svetitsky,  Phys. Rev. {\bf D 48} (1993) 190.  hep-ph/9212206;
F.Cooper, J. Dawson, Y. Kluger and H. Shepard,  Nuclear Physics
A566 (1994) 395c.
\bibitem{tr1} A. Bialas and W. Czyz, Phys. Rev. D30 (1984) 2371; A. Bialas,
W. Czyz, A. Dyrek and W. Florkowski, Nucl. Phys. B296 (1988) 611;
K. Kajantie and T. Matsui, Phys. Lett. 164B (1985) 373; G. Gatoff, A. K.
Kerman and T. Matsui, Phys. Rev. D36 (1987) 114.
\bibitem{grr1} F. Cooper, E. Mottola and G. C. Nayak, Phys. Lett. B555 (2003) 181; G. C. Nayak, A. Dumitru, L. McLerran and W. Greiner, Nucl. Phys. A687 (2001) 457; G. C. Nayak and V. Ravishankar, Phys. Rev. D55 (1997) 6877; Phys. Rev. C58 (1998) 356; R. S. Bhalerao and G. C. Nayak, Phys. Rev.  C61 (2000) 054907; G. C. Nayak, JHEP 9802:005,1998; Phys. Lett. B442 (1998) 427; G. C. Nayak and W. Greiner, hep-th/0001009; D. D. Dietrich, G. C. Nayak and W. Greiner, Phys. Rev. D64 (2001) 074006; hep-ph/0009178; J. Phys. G28 (2002) 2001; C-W. Kao, G. C. Nayak and W. Greiner, Phys. Rev. D66 (2002) 034017; F. Cooper, C-W. Kao and G. C. Nayak, Phys. Rev. D66 (2002) 114016.
\end{thebibliography}
\end{document}